\documentclass{article}

\usepackage{PRIMEarxiv}

\usepackage{amsmath, amssymb}
\usepackage{algorithm}
\usepackage{algpseudocode}
\usepackage{array} 
\usepackage[utf8]{inputenc} 
\usepackage[T1]{fontenc}    
\usepackage{hyperref}       
\usepackage{url}            
\usepackage{booktabs}       
\usepackage{amsfonts}       
\usepackage{nicefrac}       
\usepackage{microtype}      
\usepackage{lipsum}
\usepackage{fancyhdr}       
\usepackage{graphicx}       
\graphicspath{{media/}}     
\usepackage{authblk}

\usepackage{graphicx}
\usepackage{authblk}
\usepackage[most]{tcolorbox} 
\definecolor{myLightPurple}{RGB}{230, 222, 252}

\tcbset{
  titleabstractbox/.style={
    colback=myLightPurple,    
    colframe=myLightPurple,   
    boxrule=0pt,              
    rounded corners=1.5mm,    
    left=1em, right=1em,      
    top=1em, bottom=1em,      
    before=\vspace{12pt},     
    after=\vspace{12pt}       
  }
}

\newtcolorbox{abstractmodule}{
  enhanced,
  colback=white,            
  colframe=black!65,        
  boxrule=0.8pt,            
  arc=10pt,                 
  drop fuzzy shadow,        
  boxsep=0pt,
  left=8pt, right=8pt, top=8pt, bottom=8pt, 
  before=\vspace{1em},      
  after=\vspace{1em}        
}

\usepackage{fancyhdr}
\usepackage{adjustbox}

\fancypagestyle{firstpage}{
  \fancyhead{}
  \lhead{
      \adjustbox{valign=m}{\includegraphics[height=0.9cm]{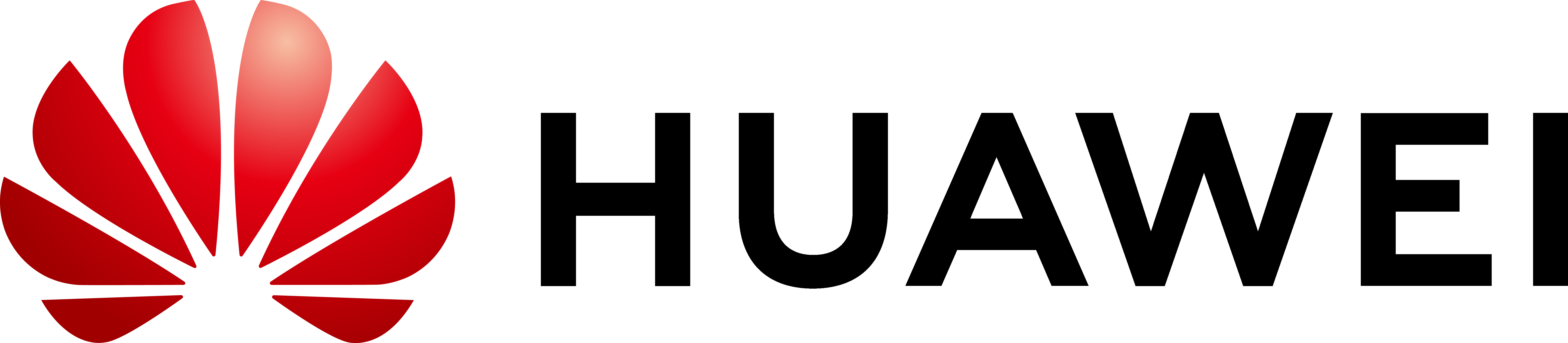}}
      \hspace{1em}
      \adjustbox{valign=m}{\includegraphics[height=1.0cm]{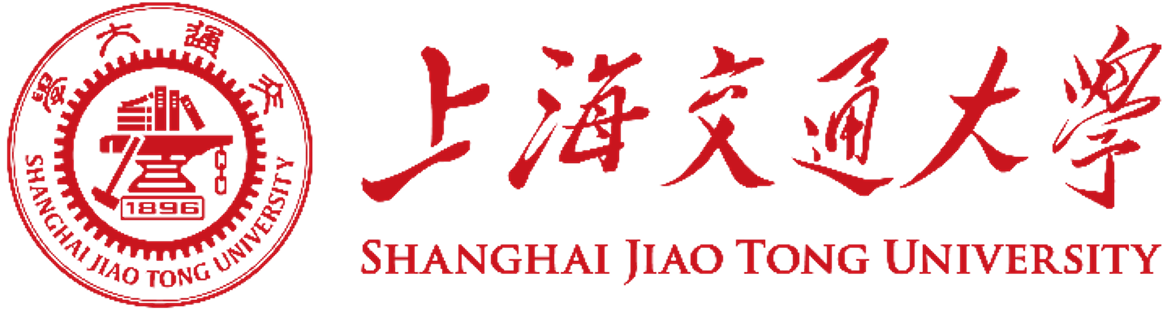}}
  }
  
  \setlength{\headsep}{1.3cm}
}

\newcommand{\archname}{HyperOffload}
\title{%
  \begin{minipage}[c]{0.1\textwidth}
    \includegraphics[height=3.2\baselineskip, keepaspectratio]{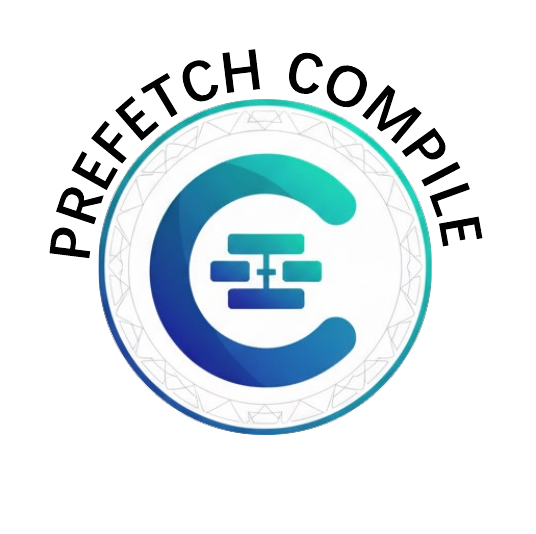}
  \end{minipage}%
  \hspace{0.8cm}%
  \begin{minipage}[c]{0.8\textwidth}
    \archname{}: Graph-Driven Hierarchical Memory  Management
    for Large Language Models
    on SuperNode Architectures
  \end{minipage}%
}

\author[1,*]{Fangxin Liu}
\author[2,*]{Qinghua Zhang}
\author[1,*]{Hanjing Shen}
\author[2]{Zhibo Liang}

\author[1,\dag]{\\Li Jiang}
\author[1]{Haibing Guan}
\author[2]{Chong Bao}
\author[2]{Xuefeng Jin}
\affil[1]{Shanghai Jiao Tong University, Shanghai, China}
\affil[2]{Huawei Technologies Co., Ltd., China}

\begin{document}

\maketitle
\thispagestyle{firstpage}

\begingroup
\renewcommand\thefootnote{*}
\footnotetext{These authors contributed equally to this work.}
\renewcommand\thefootnote{†}
\footnotetext{Corresponding author: Li Jiang (\texttt{ljiang@sjtu.edu.cn})}
\endgroup

\begin{abstract}
The rapid evolution of Large Language Models (LLMs) towards long-context reasoning and sparse architectures has pushed memory requirements far beyond the capacity of individual device HBM. While emerging supernode architectures offer terabyte-scale shared memory pools via high-bandwidth interconnects, existing software stacks fail to exploit this hardware effectively. Current runtime-based offloading and swapping techniques operate with a local view, leading to reactive scheduling and exposed communication latency that stall the computation pipeline.

In this paper, we propose the SuperNode Memory Management Framework (\textbf{\archname{}}). It employs a compiler-assisted approach that leverages graph-driven memory management to treat remote memory access as explicit operations in the computation graph, specifically designed for hierarchical SuperNode architectures. Unlike reactive runtime systems, SuperNode represents data movement using cache operators within the compiler's Intermediate Representation (IR). This design enables a global, compile-time analysis of tensor lifetimes and execution dependencies. Leveraging this visibility, we develop a global execution-order refinement algorithm that statically schedules data transfers to hide remote memory latency behind compute-intensive regions. We implement SuperNode within the production deep learning framework MindSpore, adding a remote memory backend and specialized compiler passes. Evaluation on representative LLM workloads shows that SuperNode reduces peak device memory usage by up to 26\% for inference while maintaining end-to-end performance. Our work demonstrates that integrating memory-augmented hardware into the compiler's optimization framework is essential for scaling next-generation AI workloads.
\end{abstract}


\section{Introduction}
\label{sec:intro}
As large language models (LLMs) evolve toward longer-context reasoning~\cite{liu2024lost}, multimodal capabilities~\cite{li2023blip}, and sparsely activated MoE architectures~\cite{fedus2022switch}, modern AI systems increasingly face a severe memory wall. In inference workloads such as retrieval-augmented generation (RAG)~\cite{lewis2020retrieval} and multi-turn dialogue, as well as in training workloads that demand larger batch sizes and deeper networks, the data footprint grows faster than the capacity of on-device memory. As a result, memory capacity has become a first-order constraint on both model scalability and system performance~\cite{liu2024spark}.

On the inference side, multi-level key--value (KV) caches dominate memory consumption~\cite{kwon2023efficient}. Under long-sequence workloads, KV caches alone can exceed the capacity of GPU memory. On the training side, peak activation memory frequently triggers out-of-memory (OOM) failures~\cite{chen2021actnn}. Existing techniques such as sparsification (e.g., NSA)~\cite{raihan2020sparse} and activation recomputation partially alleviate memory pressure~\cite{rajbhandari2020zero}, but they either degrade model quality or introduce significant computational overhead. These approaches therefore do not address the fundamental limitation: insufficient memory capacity. Consequently, modern systems increasingly rely on frequent data movement across heterogeneous memory tiers. In many cases, the cost of these memory transfers exceeds that of computation, making data movement the dominant performance bottleneck.

Recent advances in hardware architectures offer a new foundation for addressing this challenge. Emerging supernode architectures provide a large-capacity shared memory pool connected by high-bandwidth interconnects~\cite{li2025fenghuang, yang2025beluga}, enabling GPUs to directly access remote memory with low latency. This capability enables a memory-augmented computation paradigm, where shared memory functions as an extension of device memory for model parameters, KV caches, and activations. With sufficient bandwidth and effective scheduling, such architectures can overcome the intrinsic memory limits of a single device, supporting substantially longer context windows and larger model scales. In principle, this can significantly reduce per-token serving cost and expand the scalability envelope of both training and inference.

\begin{figure*}
    \centering
    \includegraphics[width=1\linewidth]{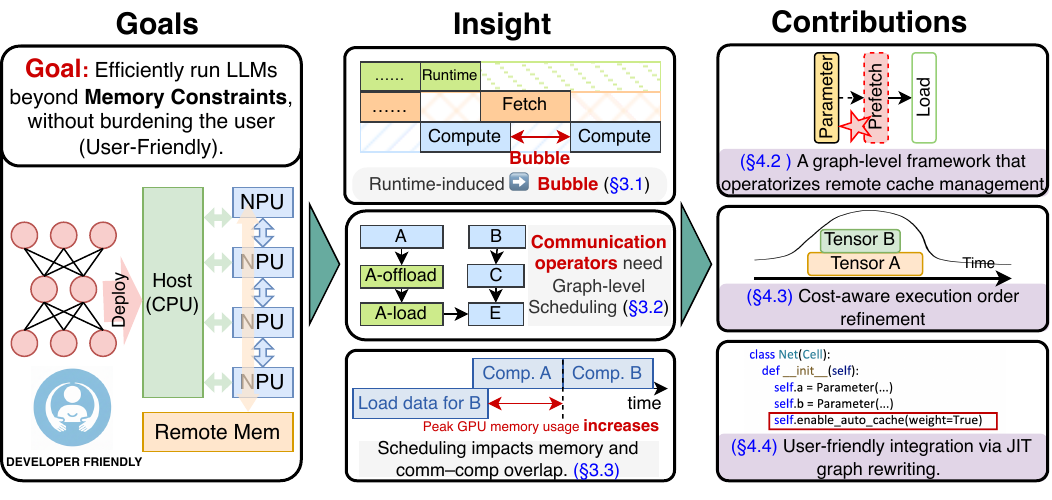}
    \caption{Motivation of graph-level remote memory scheduling. Runtime-driven data movement leads to execution bubbles and increased peak device memory usage. By explicitly scheduling communication at the graph level, we enable efficient overlap between computation and remote memory access, allowing large models to run beyond device memory limits transparently.}
    \label{fig:goals-insight-motivation}
\end{figure*}

However, fully realizing this potential requires rethinking the software stack. As illustrated in Figure~\ref{fig:goals-insight-motivation}, existing techniques such as Offload~\cite{ren2021zero}, ZeRO~\cite{rajbhandari2020zero}, pipeline parallelism~\cite{huang2019gpipe, narayanan2019pipedream}, and recomputation operate~\cite{huang2025stalloc} primarily at the runtime layer, after the computation graph has been constructed. This late-binding execution model introduces two fundamental limitations. First, the runtime lacks global visibility into the computation graph and future execution paths. Scheduling decisions are made reactively based on instantaneous memory pressure, which prevents proactive data prefetching and leads to exposed communication delays. Second, the runtime has limited control over execution order. Without graph-level planning, data transfers often contend with computation for resources, and communication latency cannot be reliably hidden behind compute-intensive operations. As a result, even with terabyte-scale shared memory and high-bandwidth links, current systems suffer from pipeline stalls and fail to approach the performance limits implied by the hardware.

Our key insight is that remote memory access must be elevated from a runtime mechanism to a first-class concept in the computation graph. As shown in Figure~\ref{fig:goals-insight-motivation}, offload and reload operations should be explicitly represented as graph operators that the compiler can analyze, schedule, and optimize. Exposing communication at the graph level enables deterministic memory planning with global knowledge of data lifetimes and execution dependencies. This design provides three major benefits. First, it enables precise compile-time analysis of memory lifecycles, eliminating redundant allocations and reducing peak memory usage. Second, it allows static reordering of operations to achieve near-complete overlap between communication and computation. Third, it integrates shared memory pools into a unified, graph-level optimization framework.

Building on these insights, we propose \textbf{\archname{}}, an AI compiler framework tailored for Supernode architectures. By integrating Graph-driven Memory management via specialized cache operators at the MindIR level, \archname{} treats remote memory access as explicit, schedulable entities. Combined with a global execution-order refinement algorithm, our framework enables efficient and deterministic remote memory extension for both training and inference. This approach fully exploits the capabilities of supernode architectures, bridging the gap between hardware potential and software execution through \archname{}'s unified memory orchestration.

In summary, our contributions are as follows:
\begin{itemize}
  \item \textbf{Framework for Remote Cache Operations.} We propose a novel framework that treats remote cache operations as first-class entities, integrating them into the global MindIR graph scheduling pipeline. This approach enables compiler-aware memory orchestration, improving overall system efficiency.
  \item \textbf{Graph-Driven Execution-Order Optimization.} We introduce Graph-Driven Execution-Order Optimization, a novel algorithm that effectively hides heterogeneous offload and reload operations at the graph level. This optimization maximizes the overlap between computation and communication, thereby minimizing peak memory usage.
  \item \textbf{Hierarchical Memory Execution Model.} We develop a unified hierarchical memory execution model that reliably supports training, inference, and ensures high availability at the cluster level. This model enhances scalability and adaptability across diverse workloads.
\end{itemize}


\section{Background}
\label{sec:backg}

\subsection{Memory Characteristics of LLM Workloads}

\paragraph{a) Inference vs. Training Memory Components}

Large language model (LLM) workloads exhibit fundamentally different memory behaviors in inference and training.

During inference, the memory footprint of LLMs is primarily dominated by model parameters and KV caches. While weights remain static across inference requests, the KV caches generated by the attention mechanism grow linearly with both context length and the number of active tokens, making them the main contributor to peak memory usage in long-context or multi-turn dialogue scenarios~\cite{liu2025aster}. For example, previous studies show that when sequence lengths scale to tens of thousands of tokens, the memory consumption of KV caches can significantly exceed that of model weights, accounting for the majority of total memory usage\cite{hooper2024kvquant}. In complex inference pipelines such as RAG, different sub-queries maintain independent KV caches, resulting in multiple concurrent KV cache lifecycles within the system and further amplifying memory demand~\cite{liu2025flexquant}.

In addition to model parameters, training workloads require retaining intermediate activations for backpropagation as well as optimizer states, leading to substantially higher memory pressure. On one hand, adaptive optimizers such as Adam maintain momentum and variance tensors for each parameter, and the total memory footprint of these optimizer states is often comparable to or even larger than that of the model parameters themselves. On the other hand, the large number of activation tensors produced during the forward pass of deep models also consumes significant memory. For instance, training a 1.5B-parameter GPT-2 model with a sequence length of 1024 and batch size 32 requires approximately 60 GB of GPU memory for forward activations alone; enabling activation checkpointing can reduce this requirement to around 8 GB \cite{rajbhandari2020zero}. Nevertheless, for larger models (e.g., 100B parameters), activation memory can still reach tens of gigabytes even with checkpointing enabled. In general, the memory footprint during training can be divided into two main components: model states (parameters, gradients, and optimizer states) and residual states (intermediate activations, temporary buffers, and memory fragmentation). Among these, model states typically account for the majority of memory usage, with the remainder consumed by activations and buffers, as illustrated by the GPT-2 example above.

\paragraph{b) Memory Constraints of LLMs}

Although the capacity of accelerator memory continues to increase, the memory requirements of LLM workloads grow at a much faster pace. For example, previous work estimates that training a trillion-parameter model using the Adam optimizer with FP16 precision requires approximately 16 TB of memory\cite{rajbhandari2020zero}, far exceeding the tens of gigabytes available on a single GPU. Moreover, long-context inference can cause the combined footprint of model weights and KV caches to exceed the memory capacity of a single device, necessitating coordination across multiple devices\cite{yuzuguler2025preserve}.

To address this challenge, existing systems often adopt heterogeneous memory hierarchies, offloading part of the data asynchronously to CPU memory or remote storage. However, such naive offloading strategies introduce significant communication overhead and latency, which are difficult to hide, particularly in bandwidth-constrained environments. Recent studies show that even high-bandwidth HBM cannot sustain the continuous access to model weights and KV caches during decoding, making inference performance limited by external memory bandwidth~\cite{yuzuguler2025preserve}. In training, offloading activation checkpoints to CPU memory can significantly reduce on-device storage, but it adds data transfer overhead~\cite{rajbhandari2020zero}. As a result, memory capacity remains a primary bottleneck for both LLM training and inference. This limitation motivates new hardware and software approaches to expand the effective memory space.


\subsection{SuperNode Architecture}

\begin{figure}
    \centering
    \includegraphics[width=1\linewidth]{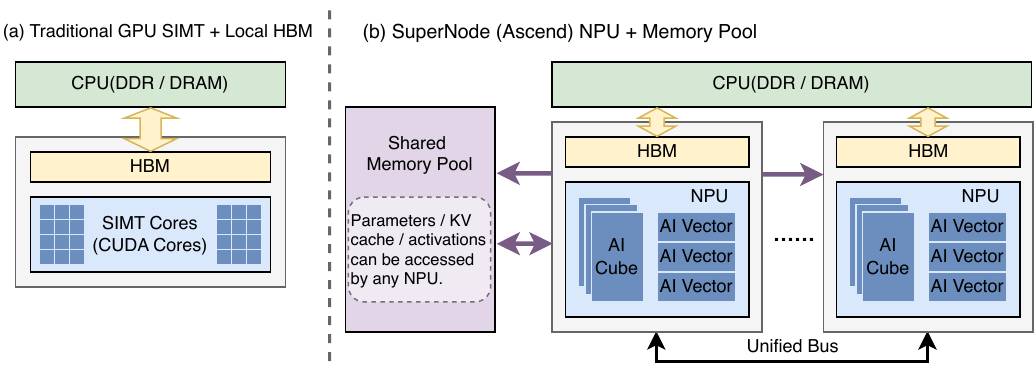}
    \caption{SuperNode Hardware Architecture. Compared to (a) traditional SIMT-based GPUs with per-device HBM, (b) SuperNode architectures integrate tile-based NPUs interconnected by ultra-high-bandwidth links and a large shared memory pool. The shared memory pool provides a unified address space, enabling symmetric access to model parameters, KV caches, and activations across all NPUs within the SuperNode.}
    \label{fig:arch}
\end{figure}

Figure~\ref{fig:arch} provides an overview of the SuperNode hardware architecture and highlights the key differences between traditional SIMT-based GPUs and tile-based NPU systems with shared memory pools.

\subsubsection{a) From SIMT GPU to Tile-Based Ascend NPU}

As illustrated in Figure~\ref{fig:arch}(a), traditional GPUs are based on the SIMT execution model, leveraging thousands of homogeneous CUDA cores together with high-bandwidth on-device HBM to maximize throughput~\cite{choquette2022nvidia}. While this design is highly effective for compute-intensive workloads, it increasingly faces scalability limitations in extremely memory-intensive scenarios such as LLMs, where model states and intermediate activations grow rapidly with sequence length.

In contrast, next-generation tile-based NPUs, such as the Huawei Ascend series shown in Figure~\ref{fig:arch}(b), adopt heterogeneous multi-core designs that integrate specialized tensor computation units and vector engines~\cite{liao2021ascend}. These units are tightly coupled with deeper and higher-bandwidth memory hierarchies, including large-capacity on-package HBM and cross-die interconnects, with aggregate bandwidths reaching the TB/s scale~\cite{zuo2025serving}.

Taking the Ascend 910 as an example, it is a dual-die packaged chip in which two compute dies share eight stacks of on-package HBM and are connected via high-bandwidth cross-die interconnects. Each die integrates 24 AI Cube matrix compute cores and 48 AI Vector cores, optimized for BF16, FP16, and INT8 computation formats~\cite{zuo2025serving}. As conceptually depicted in Figure~\ref{fig:arch}(b), such tightly integrated compute and memory subsystems improve compute density and memory bandwidth, enabling a better balance between computation and communication for large-scale AI workloads.

\subsubsection{b) Memory Pools}

A defining feature of the SuperNode architecture, as shown in Figure~\ref{fig:arch}(b), is the presence of large-capacity shared memory pools that are directly connected to multiple NPU/GPU devices through ultra-high-bandwidth interconnects. These memory pools expose a unified addressable space, effectively extending on-device memory and enabling symmetric access from all NPUs within the SuperNode.

In Huawei’s CloudMatrix384 architecture, for example, a SuperNode is constructed from 384 Ascend 910 NPUs and 192 CPUs interconnected via an ultra-high-bandwidth Unified Bus, enabling fully connected direct communication among all compute and storage components. This design allows the entire SuperNode to operate as a single logical entity, in which compute resources and memory are pooled and uniformly accessible~\cite{zuo2025serving}.

As a result, model parameters, KV caches, and activations stored in the shared memory pool can be accessed symmetrically by any NPU, regardless of where the data is originally produced or stored, with consistent bandwidth and latency characteristics. This symmetric access model, visually highlighted in Figure~\ref{fig:arch}(b), fundamentally differs from traditional host-memory offloading and provides a much stronger foundation for large models and long-context workloads.

However, fully exploiting shared memory pools requires careful software-level orchestration. Without compiler- or computation-graph-level coordination, direct access to external memory can still expose communication latency and waste bandwidth, preventing effective overlap between data movement and computation.

This challenge motivates techniques such as operatorized prefetching. Modern ML graph optimization frameworks (e.g., Huawei CANN Graph Engine) can automatically insert prefetch operators into the computation graph and execute them asynchronously to overlap data movement with computation and communication. For example, the PRESERVE framework analyzes the computation graph and inserts prefetch operators for model weights and KV caches between attention computation and communication phases, allowing data to be proactively loaded into on-chip caches during communication\cite{yuzuguler2025preserve}. Such mechanisms establish proper temporal alignment between memory transfers and computation, effectively mitigating memory bandwidth bottlenecks and providing software support for running large models efficiently on SuperNode architectures.

\section{Motivation}
\label{sec:moti}

\subsection{Limitations of Runtime-Driven Prefetching}
\begin{figure}
    \centering
    \includegraphics[width=0.5\linewidth]{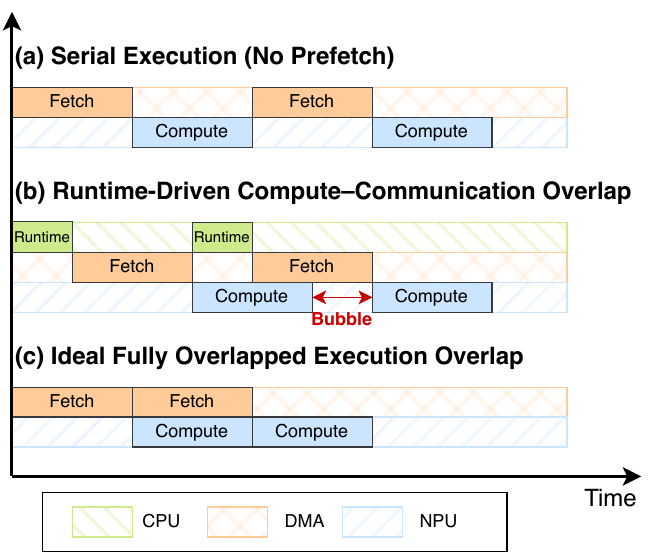}
    \caption{Comparison of execution timelines under different compute–communication orchestration strategies. (a) Serial execution without data prefetching, where NPU computation and DMA transfers are fully serialized. (b) Runtime-driven compute–communication overlap, in which the CPU orchestrates DMA prefetching during NPU execution, introducing runtime-induced bubbles. (c) Ideal fully overlapped execution, where DMA prefetching and NPU computation are statically orchestrated ahead of time, achieving bubble-free parallelism.}
    \label{fig:runtime-bubble}
\end{figure}
We evaluate  the overhead of existing prefetching mechanisms, we evaluate the LLaMA3-8B~\cite{grattafiori2024llama} model on an Ascend 910C platform (eight NPUs per node). While the baseline execution completes in 5.5 seconds, enabling runtime-driven prefetching increases the end-to-end latency to 15 seconds, representing a $2.7\times$ slowdown. A breakdown of this latency reveals that 9 seconds are spent on computation where communication is not effectively hidden, and 6.7 seconds are consumed by memory compaction and system-level management. As shown in Figure~\ref{fig:runtime-bubble}(b), the execution exhibits only partial compute-communication overlap, with frequent runtime interventions introducing significant bubbles in the device timeline.

This performance degradation stems from three structural inefficiencies inherent in runtime-driven designs. First, runtime-driven prefetching frequently interrupts the accelerator execution pipeline. Each prefetch operation requires the CPU to inspect runtime states, issue DMA requests, and synchronize with the device. These control-path overheads inject idle gaps that are difficult to overlap with computation, particularly in fine-grained or high-frequency prefetching scenarios. Second, the runtime lacks visibility into the future operator topology. Without knowledge of the upcoming computation path, the system can only trigger data transfers reactively, shortly before consumption. This prevents the system from exploiting earlier idle windows for proactive data movement, limiting the potential for latency hiding. Third, in communication-intensive workloads such as multi-level KV cache management or large-batch inference, these scheduling overheads cannot be effectively amortized. This reactive behavior contrasts with the idealized execution in Figure~\ref{fig:runtime-bubble}(c), where data movement is scheduled ahead of time to fully utilize the bandwidth of the shared memory pool.

In summary, these issues arise because the runtime lacks a global view of the computation graph and cannot reorder operators. Consequently, the shared memory pool is treated as a reactive storage target rather than a seamless extension of device memory.

\subsection{Compiler Benefits Enabled by Operatorization}
To address the limitations of runtime-driven mechanisms, we propose elevating remote memory operations to first-class operators within the computation graph. Once offload and prefetch operations are operatorized, they become visible to the compiler and can participate in graph-level analysis and optimization.
\begin{itemize}
    \item \textbf{Global Visibility of Memory Lifecycles:} Operatorization enables the compiler to gain a comprehensive view of memory lifecycles. With remote cache operations explicitly represented as graph nodes, the compiler can track when data is produced, consumed, offloaded, and reloaded. This allows for precise lifetime analysis, which is fundamentally unavailable in runtime-only approaches.

    \item \textbf{Predictable Memory Management:} In a static computation graph, memory allocation and release points are defined at compile time. This determinism reduces redundant memory demand from dynamic loading, ensuring that data is materialized only when necessary and avoiding conservative retention driven by runtime uncertainties.

    \item \textbf{Systematic Just-In-Time (JIT) Optimization:} The compiler can automatically pipeline communication operators alongside compute operators, enabling overlapping data transfers and computation without requiring users to manually insert low-level synchronization primitives, such as \textit{cudaStreamSynchronize}. This shift alleviates the performance tuning burden from users to the compiler, fostering more robust and portable optimizations.
\end{itemize}

\subsection{Challenges of Nondeterministic Execution Order in Effective Prefetching}
\begin{figure}
    \centering
    \includegraphics[width=0.5\linewidth]{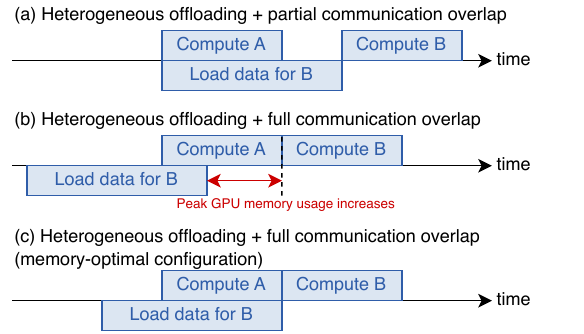}
    \caption{Comparison of Communication Overlap Strategies in Heterogeneous Offloading Systems.}
    \label{fig:communication-overlap}
\end{figure}
While operatorization exposes communication operations to the compiler and enables graph-level reasoning, it does not by itself guarantee stable or efficient prefetching. In practice, end-to-end performance is highly sensitive to the concrete execution order chosen for the computation graph. Figure~\ref{fig:communication-overlap} shows that different execution orders for the same set of compute and communication operators yield dramatically different trade-offs between communication overlap and peak memory usage.

The computation graph topology is deterministic, but the relative order among independent operators is often unspecified and left to the runtime scheduler. This nondeterminism leads to inefficient patterns. In Figure~\ref{fig:communication-overlap}(a), prefetching occurs too late: transfers do not complete before their consumers run, causing computation to stall on data and exposing communication latency. Memory usage may remain low, but throughput drops due to insufficient overlap. By contrast, Figure~\ref{fig:communication-overlap}(b) shows prefetching scheduled too early: transfers hide latency but prefetched data occupies device memory for long periods, increasing peak memory usage and waste.

These cases demonstrate a significant trade-off between memory efficiency and performance: prefetching must be early enough to avoid stalls yet late enough to avoid unnecessary residency. Without explicit control over execution order, systems cannot reliably attain both memory-optimal and performance-optimal configurations, which results in variable performance and unpredictable peak memory across runs.

To achieve effective memory reduction and reliable communication-computation overlap, explicit Graph-Driven Execution-Order Optimization is essential. As shown in Figure~\ref{fig:communication-overlap}(c), an optimized execution sequence that aligns data loading with computation can fully hide communication latency while preventing unnecessary memory residency. This optimization must take into account operator execution times, data transfer volumes, and available bandwidth, ensuring that data loading occurs at the optimal moment.  Only through execution-order optimization can operatorized prefetching fully realize its potential, minimizing peak memory usage while maximizing overlap efficiency.

\section{\archname{} Framework: Compiler-Assisted Design and Optimization}

\subsection{Framework Overview}
The \archname{} Framework is a compiler-assisted execution system designed to fully exploit large-capacity shared memory pools on SuperNode architectures. The key insight is to elevate remote memory accesses from opaque runtime behaviors to explicit, schedulable entities at the computation graph level, enabling deterministic memory planning and effective communication–computation overlap. Figure~xxx illustrates the framework's overall architecture. Starting with a conventional MindSpore computation graph, the \archname{} Framework transforms it into a memory-aware MindIR graph that jointly orchestrates computation and remote memory access.


\textbf{Graph-Level Optimization for Remote Memory Access.} The key insight of framework is its focus on graph-level coordination of remote memory. Rather than depending on runtime-driven offloading and on-demand loading, the \archname{} Framework explicitly models remote memory interactions within MindIR. Operations related to remote caching, such as prefetching data into device memory, persisting tensors back to remote memory, and releasing device residency, are treated as first-class operators within the graph. This design enables the compiler to globally analyze tensor lifetimes, memory residency, and data movement. Consequently, memory transfers transition from being implicit side effects to visible components that can actively participate in dependency analysis and scheduling decisions.

\textbf{Latency Hiding via Graph-Driven Execution-Order Optimization.} Making cache operations explicit allows for additional execution-level optimizations. In MindIR, operators without data dependencies can execute in any order, which may lead to unstable overlap behavior: prefetches might occur too early, unnecessarily occupying device memory, or too late, resulting in computation stalls. The \archname{} Framework tackles this challenge by optimizing the execution order of independent operators based on a cost model considering operator execution time, data transfer volume, and interconnect bandwidth. The compiler schedules prefetch operations just in time—early enough to complete before their use, yet late enough to minimize memory footprint—ensuring that remote memory latency is consistently hidden behind computation.

\textbf{Non-intrusive Integration via JIT Graph Rewriting.} Importantly, the \archname{} Framework is designed to be non-intrusive for users. Models are developed using standard MindSpore APIs without explicit offload or prefetch annotations. All cache operator insertions and execution order optimizations occur automatically during JIT compilation through graph rewriting. This approach maintains user productivity while allowing advanced users the option to fine-tune memory behavior through specialized interfaces. From the user’s perspective, the framework seamlessly transforms shared remote memory into an effective extension of device memory, all without introducing artificial data dependencies, thus preserving the optimization flexibility of the original graph.

\subsection{Operatorized Remote Cache Management in MindIR}
A fundamental challenge in hierarchical memory systems is that remote memory accesses are typically managed as implicit runtime behaviors. Such accesses remain invisible to the compiler, inhibiting global reasoning about tensor lifetimes, data movement, and execution overlap. The \archname{} Framework addresses this limitation by operatorizing remote cache management, making it an explicit component of the MindIR computation graph. By modeling remote memory interactions as first-class operators, the framework facilitates compile-time analysis and scheduling of memory movement, laying the groundwork for deterministic latency hiding.

\subsubsection{Cache Operators as First-Class Graph Nodes}

The \archname{} Framework introduces a suite of cache-related operators into MindIR, including \textit{Prefetch}, \textit{Store}, and \textit{Detach}. These operators explicitly represent remote-to-device loading, device-to-remote eviction, and residency control, respectively. Unlike runtime-triggered memory operations, cache operators are directly embedded into the computation graph and actively participate in standard graph analyses, such as dependency inference and topological ordering. Each cache operator has well-defined semantics: a \textit{Prefetch} operator asynchronously initiates data transfer from remote memory to device memory, ensuring correctness by enforcing completion before the first consumer; a \textit{Store} operator safely releases device memory by transferring data back to remote storage; and a \textit{Detach} operator marks tensors whose device residency can be relinquished. This operatorization transforms memory movement from an opaque side effect into a schedulable unit, enabling the compiler to reason jointly about computation and communication.

\subsubsection{Compile-Time Prefetch Insertion}

Once cache operations are explicitly represented, the compiler can determine when and where to insert them. During graph compilation, \archname{} Framework analyzes tensor lifetimes and future access patterns to identify data that will be required by upcoming operators but need not remain resident in device memory continuously. Based on this analysis, \textit{Prefetch} operators are automatically inserted ahead of their first use, allowing data to be transferred asynchronously from remote memory before it is consumed. Because prefetching is performed at compile time, the framework can plan memory movement across operator boundaries rather than reacting to runtime memory pressure. This design eliminates the need for conservative, on-demand loading strategies that often expose communication latency on the critical path.

\subsection{Graph-Driven Execution-Order Optimization for Latency Hiding}

Although operatorized remote cache management enables asynchronous data movement, its effectiveness critically depends on when cache operators are executed. In MindIR, the execution order among operators without explicit data dependencies is not fixed and may vary across runs. Such nondeterminism can significantly impact both memory footprint and performance, especially when remote memory operations are involved. \archname{} Framework addresses this challenge through Graph-Driven Execution-Order Optimization, a compiler-driven optimization that refines the relative order of independent operators based on graph-level analysis. This optimization achieves just-in-time prefetching and maximizes communication--computation overlap, without introducing artificial dependencies or modifying the original computation graph.

Algorithm~\ref{alg:execution-order} summarizes the compiler procedure used in Graph-Driven Execution-Order Optimization. Starting from a valid topological order, the algorithm evaluates multiple feasible positions for each cache operator using a cost model that captures both memory residency duration and exposed communication latency, and selects a placement that best aligns data availability with its first use.

\begin{algorithm}[t]
    \caption{Graph-Driven Execution-Order Optimization}
    \label{alg:execution-order}
    \begin{algorithmic}[1]
        \Require Computation graph $G=(V,E)$ with cache operators
        \Require Execution cost $C_\text{comp}(v)$ for $v \in V$
        \Require Transfer cost $C_\text{trans}(c)$ for cache operator $c$
        \Ensure Refined execution order $\mathcal{O}$
        
        \State $\mathcal{O} \gets \text{topo}(G)$
        \State $\mathcal{C} \gets \{ c \in \mathcal{O} \mid c \text{ is independent cache operator} \}$
        
        \For{each $c \in \mathcal{C}$}
            \State $u \gets$ first consumer of $c$
            \State $\text{Pos}_c \gets$ feasible positions of $c$ in $\mathcal{O}$
            
            \For{each $p \in \text{Pos}_c$}
                \State $T_\text{trans}(c,p) \gets$ transfer completion time at $p$
                \State $L_\text{overlap}(c,p) \gets$ overlap with computation before $u$
                \State $C(p) \gets$ cost function based on latency and memory
            \EndFor
        
            \State $p^* \gets \arg\min_{p \in \text{Pos}_c} C(p)$
            \State $\mathcal{O} \gets \mathcal{O}[c \to p^*]$
        \EndFor
        
        \State \Return $\mathcal{O}$
    \end{algorithmic}
\end{algorithm}

\subsection{Non-Intrusive Integration}
\begin{figure}
    \centering
    \includegraphics[width=1\linewidth]{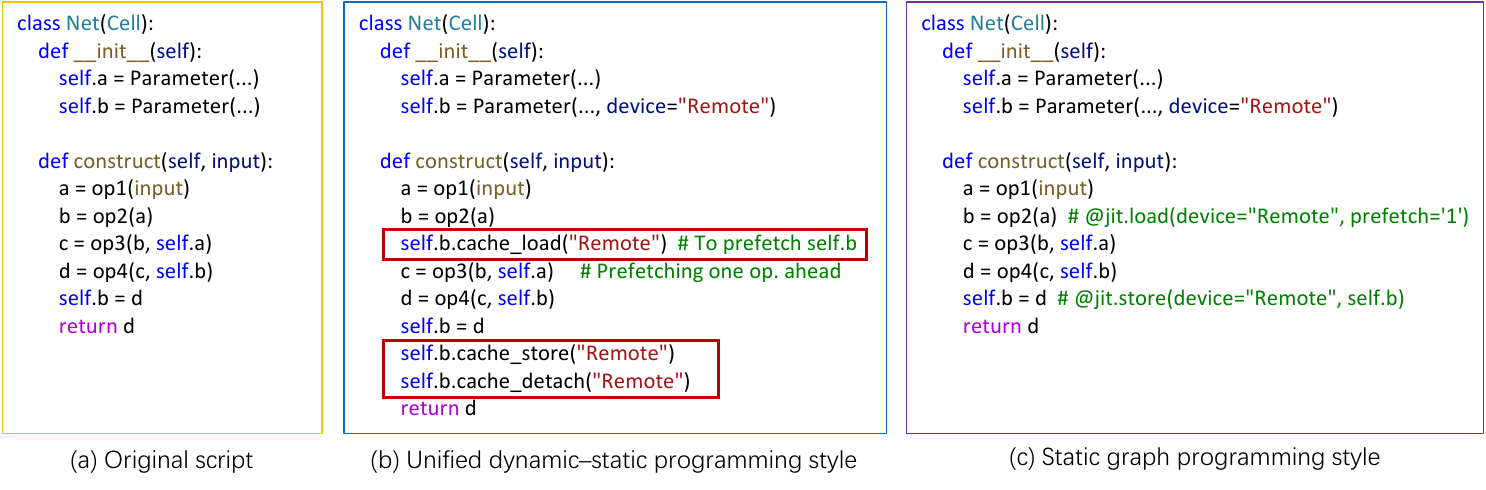}
    \caption{Hierarchical memory interface usage example. The figure demonstrates how model parameters are placed in remote memory and how cache prefetch, store, and detach operations are exposed to users. Both explicit API calls and JIT-based static graph annotations are shown, illustrating flexible and unified interfaces for hierarchical memory management.}
    \label{fig:code}
\end{figure}
A main design goal of the \archname{} Framework is to add hierarchical memory management without adding programming effort for users. The framework integrates into existing MindSpore workflows so that users can keep their current scripts and preserve both productivity and model portability.

Figure~\ref{fig:code} shows three ways to use hierarchical memory in MindSpore.

As shown in Figure~\ref{fig:code}(a), a standard MindSpore model can run with unmodified user scripts. Users do not write any code for memory placement, offloading, or prefetching. In this default mode, they only see the usual MindSpore abstractions, and hierarchical memory is completely transparent at the programming level.

In this automatic mode, hierarchical memory operations such as cache prefetching, offloading, and device detachment are inferred and inserted during JIT compilation. The compiler analyzes the computation graph, tensor lifetimes, and execution dependencies, then inserts cache operators and chooses their execution order. From the user’s point of view, remote memory behaves like an extension of device memory, and model definitions as well as training and inference scripts remain unchanged.

For workloads that need more control, SuperNode also provides optional interfaces for expert users, as shown in Figure~\ref{fig:code}(b). Through explicit cache management APIs, expert users can mark selected parameters to reside in remote memory and insert cache \textit{Load}, \textit{Store}, or \textit{Detach} operations by hand. These interfaces let domain experts tune prefetch distance, keep critical tensors in device memory, and adjust cache behavior in performance-sensitive regions, while preserving the original computation semantics.

To balance flexibility and ease of programming, the framework further supports a unified programming model based on JIT annotations, shown in Figure~\ref{fig:code}(c). In this mode, users attach high-level memory hints to computation nodes, and the compiler turns these hints into static graph transformations and cache operators. This model unifies dynamic and static graph programming styles, and it allows expert tuning without direct manipulation of low-level memory operations.

Overall, this non-intrusive design lets users adopt hierarchical memory in a gradual and safe way in production environments. General users gain higher memory efficiency and stable performance without rewriting models. Expert users can still tune memory behavior when needed. By separating memory management from model logic, the \archname{} Framework provides both ease of use and room for optimization.
\section{Case Study}

This section presents two case studies to demonstrate how the \archname{} Framework improves memory efficiency and performance on practical workloads. These memory-intensive scenarios show the limitations of runtime-based memory management and the benefits of our graph-based approach.

\subsection{Case 1: Training Scenario}
Training workloads show complex memory usage patterns across different execution phases. Beyond model parameters, a large portion of device memory is used by forward activations and optimizer states. These tensors have different lifetimes and access patterns, which makes them good targets for hierarchical memory management.


\textbf{Activation Tensors in Forward and Backward Propagation.}
During forward propagation, activation tensors are produced one by one and kept in memory until they are needed by the corresponding backward operators. In deep networks, especially those with large hidden dimensions or long sequences, these activations account for most of the peak device memory usage. In conventional runtime-driven systems, activation tensors stay in device memory until backward computation starts, or they are offloaded only when memory runs out. These reactive approaches lack global knowledge of future access, so data transfers often happen too late and stall backward computation.

The SuperNode AI Framework analyzes activation tensor lifetimes at compile time and uses cache operators in the computation graph to control their movement. Activations that are not needed immediately are offloaded to remote memory using Store operators. Prefetch operators are inserted before the backward operators that will consume them. Note that activations with very short lifetimes or fine-grained access patterns are not good candidates for remote caching, because transfer overhead can outweigh the memory savings. The scheduling algorithm in Algorithm~\ref{alg:execution-order} detects such cases at compile time and avoids offloading them. Execution order optimization schedules prefetches inside computation-heavy regions of the backward pass, so data transfer finishes before the data is used.
Such an approach reduces peak activation memory by tens of percent for representative transformer training workloads, while end-to-end iteration time stays the same.

\textbf{Optimizer State Management.}
Optimizer states, such as momentum and variance buffers, present a different problem. These tensors persist across training iterations, but they are accessed only during parameter update phases and sit idle for most of the iteration. Traditional training pipelines keep optimizer states in device memory to avoid repeated transfer overhead. This leads to high steady-state memory consumption, especially for large models with multiple optimizer states per parameter.

SuperNode treats optimizer states as long-lived but infrequently accessed tensors. After each update phase, optimizer states are offloaded to remote memory and prefetched back before the next update. Because parameter updates are usually preceded by a computation-heavy backward pass, execution order orchestration allows optimizer state prefetching to overlap fully with gradient computation.

As a result, optimizer states contribute very little to peak device memory usage while preserving update throughput. However, when update phases are short or mixed with light computation, the framework may choose to keep optimizer states in device memory to avoid unnecessary transfers.

\subsection{Case 2: Inference Scenario}

Inference workloads, especially for LLMs, show different memory patterns compared to training. Instead of temporary activations, memory usage is driven by the accumulation and reuse of KV cache tensors during autoregressive decoding.

\textbf{KV Cache Offloading in Autoregressive Inference.}
In autoregressive inference, KV caches grow as the sequence length or batch size increases. These caches can quickly exceed device memory capacity, which limits throughput or prevents the model from handling long sequences.

Current systems often manage KV caches using runtime eviction. When memory is full, the system swaps blocks to remote memory. This approach has two main problems. First, eviction and reload decisions are made locally without a global view of future access. Second, reloading KV cache blocks often happens on the critical path of token generation, which exposes the high latency of remote memory.

The SuperNode AI Framework solves these issues by representing KV cache movement as explicit operators in the computation graph. Because KV cache access across decoding steps is highly regular, the compiler can predict future usage and insert Prefetch operators before the attention computation. The framework then schedules these transfers to overlap with other operations in the decoding pipeline, such as embedding lookups and projections.

Through this graph-driven execution-order optimization, KV cache offloading becomes proactive rather than reactive. Remote memory latency is effectively hidden. This design allows the system to support longer context lengths and larger batch sizes while maintaining high inference throughput.



\section{Implementation}
The proposed \archname{} Framework integrates hierarchical memory management into MindSpore~\cite{huawei2022deep} by extending its memory abstraction, operator interfaces, and compilation flow. By operatorizing heterogeneous memory operations and exposing them via automatic and explicit mechanisms, this design enables graph-level memory optimization while strictly preserving compatibility with existing execution semantics.

\textbf{Remote Memory Management.}
The implementation introduces a remote memory pool logically distinct from host memory and device HBM. This pool is directly accessible by NPUs via DMA-capable links and acts as persistent storage for large states, including weights, optimizer states, gradients, and activations. Crucially, introducing remote memory allows for direct Remote-to-Device (R2D) and Device-to-Remote (D2R) transfers, bypassing necessary data staging through the host.

\textbf{Unified Memory Primitives.}
To manage the extended hierarchy, core memory operations are abstracted into a set of explicit data movement primitives: Host-to-Remote (H2R), Remote-to-Host (R2H), R2D, D2R, and Device-to-Device (D2D) transfers. These primitives are lowered to asynchronous DMA operations and are synchronized using MindSpore’s native stream model. Collective operations, such as AllReduce, are adapted to aggregate device-local updates before committing results back to the remote pool.

\textbf{Cache Operators in MindIR.}
Cache-related interactions are exposed as native operators within the MindIR graph representation. Operations like cache loading, storing, and detachment are compiled into graph nodes and executed using the unified memory primitives. This design ensures that remote memory scheduling and synchronization are managed within the existing graph execution framework, eliminating the need for special-case runtime logic.

\textbf{User Programming Model.}
The framework supports both explicit and implicit usage. In the explicit mode, expert users can annotate parameters or tensors for remote residency and invoke cache operations programmatically. These annotations are captured during graph construction, resulting in corresponding cache operators in the final compiled graph, facilitating fine-grained control and debugging.

For standard workloads, the automatic mode requires zero modification to user scripts. By enabling hierarchical memory support at the graph level, the compiler automatically identifies eligible tensors (e.g., large model weights) and inserts necessary load/store operations during static compilation. This grants existing models transparent access to remote memory benefits without altering training or inference logic.

Overall, the implementation demonstrates that extending existing compiler and operator abstractions is sufficient to integrate hierarchical memory into a production deep learning framework. By aligning memory operations directly with MindSpore’s graph execution model, the \archname{} Framework achieves practical deployability on SuperNode systems while maintaining a clean and extensible software architecture.

\section{Experiment}

\subsection{Experimental Methodology}

\textbf{Platform.} The experiments are conducted on the Ascend 910C NPU platform using a single-node multi-card configuration, with a typical setup of one node equipped with eight NPUs. Both training and inference are performed in MindSpore static graph mode, with hierarchical memory–related compilation and runtime options enabled. In different experiments, varying D2H/H2D bandwidth conditions are emulated through configuration to analyze the impact of communication bandwidth on overall performance.

\textbf{Networks and Tasks.}
We evaluate the proposed approach using representative large-scale models, covering both training and inference scenarios: a) Training: LLaMA-8B~\cite{grattafiori2024llama} and DeepSeek-V3~\cite{liu2024deepseek} models. b) Inference: DeepSeek-V3 integrated with the NSA (sparse attention) algorithm, with a focus on KV cache offloading and long-sequence capability.

\textbf{Baseline.}
All baselines are implemented using the native MindSpore framework without enabling hierarchical memory or execution-order optimization: a) In the training scenario, baseline models rely on parallelism strategies and activation recomputation to satisfy memory constraints, without heterogeneous offloading or communication–computation overlap. b) In the inference scenario, all KV caches reside entirely in device memory, and remote memory pools are disabled. c) In the high-availability scenario, traditional checkpoint-based mechanisms are used for model backup and recovery.

\subsection{Experimental Result for Training Scenarios}

The training experiments evaluate the performance and memory benefits of hierarchical memory in large-scale model training. Our goal is to significantly reduce end-to-end training latency, while achieving additional performance gains when sufficient Host–Device bandwidth is available. We analyze the effectiveness of the proposed approach by examining performance trends under varying bandwidth conditions and comparing end-to-end training latency across different models.

\begin{table}[htbp]
    \centering
    \caption{LLaMA-8B Training Baseline Configurations and Performance}
    \label{tab:training_llama}
    \begin{tabular}{c c c c c c}
        \hline
        Config & DP/TP/PP & Batch & GBS & Recomp. & End-to-end Cost \\ 
        \hline
        No.1 & 8/1/1 & 2 & 16 & On & 8000\,ms+ \\
        No.2 & 2/2/2 & 1 & 16 & Off & 5200\,ms \\
        \hline
    \end{tabular}
\end{table}

\begin{table}[htbp]
    \centering
    \caption{DeepSeek-V3 Tranining Baseline Configuration and Performance}
    \label{tab:training_deepseek}
    \begin{tabular}{c c c c c}
        \hline
        DP/TP/PP/EP & Batch Size & GBS & Recomp. & End-to-end Cost \\ 
        \hline
        2/2/2/4 & 1 & 16 & Disabled & 2500\,ms \\
        \hline
    \end{tabular}
\end{table}

\begin{figure}
    \centering
    \includegraphics[width=0.8\linewidth]{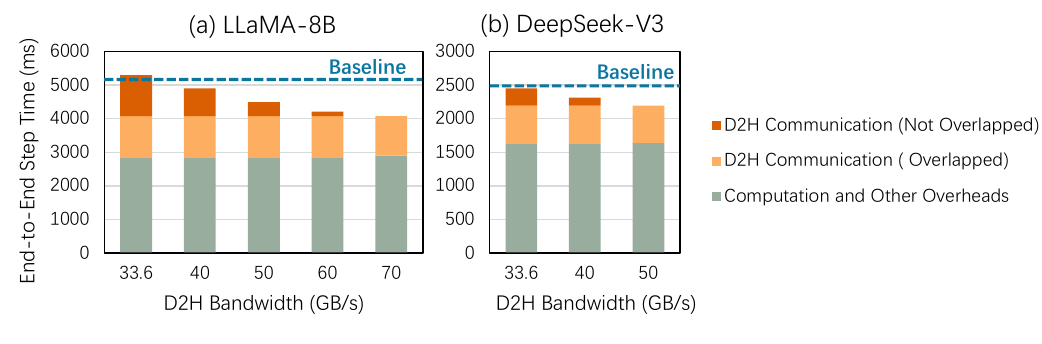}
    \caption{End-to-end training step time breakdown under different D2H bandwidths. The stacked bars decompose the per-step execution time into exposed D2H communication, overlapped D2H communication, and computation/other overheads. Results are shown for (a) LLaMA-8B~\cite{grattafiori2024llama} and (b) DeepSeek-V3~\cite{liu2024deepseek}. As the D2H bandwidth increases, exposed communication is progressively eliminated through execution-order optimization.}
    \label{fig:training}
\end{figure}

\subsubsection{LLaMA-8B Training Performance}

We first evaluate the training performance of LLaMA-8B under baseline configurations, as summarized in Table~\ref{tab:training_llama}. Due to limited device memory capacity, baseline Config No.1 frequently triggers memory defragmentation during training, resulting in substantially increased end-to-end step time. Therefore, all subsequent comparisons are conducted against baseline Config No.2, which exhibits more stable performance.

After enabling hierarchical memory, LLaMA-8B is trained with a data parallelism / tensor parallelism / pipeline parallelism configuration of 8/1/1, a batch size of 2, and a global batch size (GBS) of 16. Figure~\ref{fig:training}(a) shows the end-to-end step time breakdown and relative performance of LLaMA-8B under different D2H bandwidths. By offloading activations and a subset of parameters to remote memory and applying execution-order optimization to overlap communication with computation, the hierarchical memory approach significantly reduces end-to-end training latency.

Under the measured D2H bandwidth of 33.6 GB/s, the hierarchical memory configuration achieves performance comparable to the baseline. As the bandwidth increases to the range of 40–70 GB/s, the end-to-end training performance improves by approximately 5.7\%–21.5\% over the baseline. These results indicate that execution-order optimization effectively prevents performance degradation under bandwidth-constrained conditions, while additional bandwidth enables further hiding of communication overhead and unlocks greater performance improvements.

\subsubsection{DeepSeek-V3 Training Performance}

We then evaluate the training performance of the larger DeepSeek-V3 model. Its baseline configuration and performance are reported in Table~\ref{tab:training_deepseek}, with an end-to-end step time of approximately 2500 ms.

With hierarchical memory enabled, DeepSeek-V3 is trained using a DP / TP / PP / EP configuration of 8/1/1/4, a batch size of 2, and a global batch size of 16. Figure~\ref{fig:training}(b) presents the end-to-end training performance of DeepSeek-V3 under varying D2H bandwidths. Compared to LLaMA-8B, DeepSeek-V3 has a larger parameter count and higher computational intensity. In this setting, hierarchical memory consistently delivers stable and scalable performance benefits, reducing end-to-end training latency by approximately 2\%–12.3\% across different bandwidth configurations. Owing to the higher compute density, communication overhead is more easily hidden by computation, enabling noticeable performance gains even under moderate bandwidth conditions.

Overall, the training results demonstrate that hierarchical memory combined with execution-order optimization effectively alleviates device memory pressure during large-model training while preserving numerical correctness and execution stability. Moreover, the proposed approach significantly improves the robustness of training performance to Host–Device bandwidth variations, allowing the system to scale efficiently across diverse hardware interconnect capabilities.

\subsection{Experimental Result for Inference Scenarios}

The inference experiments focus on evaluating the impact of hierarchical memory on KV cache management and long-context inference capability. We primarily examine GPU memory footprint, the maximum supported sequence length, and end-to-end inference latency.

\subsubsection{KV Cache Offloading and Maximum Sequence Length}

 In the DeepSeek-V3 + NSA setting, offloading the entire KV cache to remote memory reduces peak device memory usage by approximately 26\%, which closely matches the KV cache size itself. As a result, under the same hardware configuration, the maximum supported sequence length increases from around 71k tokens to 123k tokens, substantially enhancing long-context inference capability. As summarized in Table~\ref{tab:kv_cache_infer}, the reduction in peak device memory closely corresponds to the KV cache size, confirming that KV cache offloading is the primary contributor to memory savings and long-context capability expansion.

\begin{table}[htbp]
    \centering
    \caption{Effect of KV Cache Offloading on Memory Footprint and Maximum Sequence Length}
    \label{tab:kv_cache_infer}
    \begin{tabular}{p{0.3\linewidth}p{0.15\linewidth}p{0.23\linewidth}>{\raggedright\arraybackslash}p{0.15\linewidth}}
        \toprule
        \textbf{Configuration} & \textbf{Baseline}& \textbf{Hierarchical Memory}&\textbf{Relative Change}\\
        \midrule
        Peak Device Memory (GB)& 61.2 & 45.0 &$\sim$-26\%\\
        Max Sequence Length (tokens)& 71k& 123k  &$\sim$1.73$\times$\\ 
        \bottomrule
    \end{tabular}
\end{table}

\subsubsection{Long-Sequence Performance Analysis}

In long-sequence inference scenarios where device memory utilization approaches its capacity limit, the baseline implementation—where the KV cache fully resides in device memory—frequently triggers memory defragmentation, leading to significant performance degradation during the prefill phase. In contrast, the hierarchical memory approach eliminates the conditions that cause defragmentation by offloading the KV cache to remote memory. As shown in Table~\ref{tab:long_seq_infer}, the number of defragmentation events is reduced from dozens to zero, resulting in a prefill latency reduction of approximately 23\% and an overall end-to-end inference latency reduction of about 13.8\%. These results indicate that, under severe memory pressure in long-sequence inference, hierarchical memory substantially improves system stability and effectively avoids performance degradation induced by memory management overheads.

\begin{table}[htbp]
    \centering
    \caption{Performance and Stability Comparison in Long-Sequence Inference}
    \label{tab:long_seq_infer}
    \begin{tabular}{p{0.3\linewidth}c >{\centering\arraybackslash}p{0.25\linewidth}c}\toprule
        
        \textbf{Metric} & \textbf{Baseline} & \textbf{Hierarchical Memory} & \textbf{Change} \\\midrule
        
        Defragmentation Events & 57 & 0 & Eliminated \\
        Prefill Latency & 129.33\,s & 99.41\,s & $-23.13\%$ \\
        End-to-End Latency & 187.21\,s & 161.41\,s & $-13.78\%$ \\ \bottomrule
        
    \end{tabular}
\end{table}

\subsubsection{Short-Sequence Performance Analysis}

For typical short-sequence inference workloads with relatively low memory pressure, the baseline and hierarchical memory approaches exhibit comparable performance during the prefill stage, indicating that KV cache offloading does not introduce noticeable overhead along the forward computation path. As shown in Table~\ref{tab:short_seq_infer}, the prefill latency differs by less than 1\% between the two configurations. However, during the decode stage, the hierarchical memory scheme incurs a more pronounced performance slowdown when sparse block granularity is increased. Specifically, the decode latency increases from 0.117\,s to 0.146\,s, corresponding to a relative slowdown of approximately 25.5\%. This overhead primarily stems from CPU-side execution of partial KV cache updates and sparse block processing. Despite the increased decode latency, the impact on end-to-end inference performance remains minimal: the overall end-to-end latency differs by only 0.15\% between the baseline and hierarchical memory configurations. This result suggests that, even under unfavorable block-size settings, the additional decode overhead accounts for only a small fraction of the total inference time and is therefore acceptable given the substantial memory savings and long-context capability gains enabled by hierarchical memory.

\begin{table}[htbp]
    \centering
    \caption{Inference Latency Breakdown in Short-Sequence Scenarios}
    \label{tab:short_seq_infer}
    \begin{tabular}{lccc}
        \hline
        \textbf{Stage} & \textbf{Baseline} & \textbf{Hierarchical Memory} & \textbf{Relative Change} \\
        \hline
        Prefill Latency (s) & 62.19 & 62.49 & -0.48\% \\
        Decode Latency (s) & 0.117 & 0.146 & -25.47\% \\
        End-to-End Latency & 177.373 & 177.109 & 0.15\% \\
        \hline
    \end{tabular}
\end{table}

\subsection{Sensitivity to Sparse Block Granularity}
When the size of sparse blocks (e.g., selection or sliding blocks) increases significantly, CPU computation and memory copy overheads during the decode stage rise noticeably, leading to further performance degradation. This observation suggests that inference performance under hierarchical memory is closely tied to the granularity of sparse structures. Future optimizations may further mitigate this overhead through data parallelism or finer-grained scheduling strategies. 

\begin{table}[]
    \centering
    \caption {Inference Latency Breakdown in Sparse Block Scenarios}
    \begin{tabular}{lccc}
        \hline
        \textbf{Metric} & \textbf{Baseline} & \textbf{Hierarchical Memory} & \textbf{Relative Change} \\
        \hline
        Peak Memory Usage & 58428M & 45828M & 21.57\% \\
        Prefill Predict Time (s) & 120.098 & 115.186 & 4.09\% \\
        Decode Predict Time (s) & 0.117 & 0.146 & -25.47\% \\
        Total Time (s) & 177.373 & 177.109 & 0.15\% \\
        \hline
    \end{tabular}
\end{table}

\section{Related Work}

Hierarchical memory management and memory optimization for large-scale deep learning workloads have been extensively studied across both training and inference scenarios~\cite{hu2025hiagent,ren2024mtm}. Existing approaches can be broadly categorized into runtime-driven memory offloading~\cite{githubGitHubAidynamodynamo}, graph-level optimization frameworks~\cite{sabne2020xla}, and KV cache–specific techniques for large language models~\cite{cai2024pyramidkv}.

\textbf{Runtime-driven memory offloading and parallelism.}
A large body of prior work focuses on reducing device memory pressure through runtime mechanisms such as parameter and activation offloading, recomputation, and distributed parallelism. Techniques including ZeRO~\cite{rajbhandari2020zero}, activation checkpointing~\cite{chen2016training, narayanan2021efficient}, pipeline parallelism~\cite{narayanan2019pipedream}, and hybrid parallel strategies~\cite{shoeybi2019megatron} aim to trade communication or computation overhead for reduced memory footprint. These methods are effective in practice but fundamentally operate at the runtime level, relying on reactive decisions based on instantaneous memory usage. As a result, they lack global visibility into future computation and data access patterns, often leading to suboptimal communication overlap, memory fragmentation, and unstable performance under tight memory constraints.

\textbf{Graph-level and compiler-based memory optimization.}
Recent research has explored compiler-assisted memory planning, where tensor lifetimes and memory allocation are analyzed statically to reduce peak memory usage~\cite{chen2016training,shah2020memory}. Systems such as static memory planners and graph schedulers leverage compile-time information to reuse buffers, reorder operators, or insert recomputation points~\cite{chen2016training}. While these approaches provide stronger determinism than runtime-only methods, most existing designs assume a single-level device memory model and do not explicitly incorporate heterogeneous or remote memory tiers into the computation graph. Consequently, communication with remote memory remains opaque to the compiler and cannot be systematically optimized or overlapped with computation.

\textbf{KV cache optimization for LLM inference.}
For inference workloads, especially large language models, a growing line of work targets the KV cache bottleneck. Techniques such as grouped-query attention (GQA)~\cite{ainslie2023gqa}, multi-query attention (MQA)~\cite{shazeer2019fast}, KV cache compression~\cite{wang2025million}, eviction policies (e.g., H2O)~\cite{zhang2023h2o}, and sparse attention mechanisms (e.g., NSA) aim to reduce KV cache size or attention computation cost. These methods primarily focus on reducing memory capacity requirements or compute complexity, but they typically assume that KV caches reside in device memory or are managed by runtime eviction policies. When offloading is employed, KV cache transfers are often triggered on demand and lie directly on the critical path of decoding, exposing remote memory latency.

\textbf{Comparison with our approach.}
In contrast, \archname{} explicitly incorporates remote memory operations as operators within the computation graph. By representing cache load, store, and detach actions as graph nodes, the framework enables global lifetime analysis and deterministic memory planning. This approach allows for the systematic overlap of communication and computation through execution-order refinement.

Unlike techniques that reduce KV cache size or rely on reactive eviction policies, our approach focuses on the precise scheduling of model states and intermediate tensors across memory tiers. Consequently, remote memory accesses are proactively planned and overlapped with computation, rather than being triggered by runtime memory pressure. Furthermore, \archname{} is orthogonal to model compression methods~\cite{liu2021improving,liu2024spark}. While compression reduces the data footprint, our framework focuses on optimizing hierarchical memory interactions at the graph level.

\section{Conclusion}

This paper presents \textbf{\archname{}}, a compiler-assisted hierarchical memory management framework that elevates remote memory operations to first-class operators in the computation graph. By operatorizing cache load and store actions and refining execution order at compile time, the framework enables deterministic memory planning and effective overlap of communication and computation. Experimental results show that the proposed approach substantially reduces device memory consumption, improves long-context inference capability, and delivers stable performance across training and inference workloads without requiring changes to user models. These results demonstrate that graph-level, compiler-aware memory management is a practical and effective foundation for fully exploiting emerging supernode architectures.

\bibliographystyle{unsrt}
\bibliography{ref}

\end{document}